\documentclass[5p,twocolumn,fleqn]{elsarticle}
\usepackage{graphicx}
\usepackage{amsmath}

\newcommand{\sub}[1]{_\mathrm{#1}}
\newcommand{\rmd}{\mathrm{d}}

\newcommand{\be}{\begin{equation}}
\newcommand{\ee}{\end{equation}}

\title{Investigating the thermodynamics of small biosystems with optical tweezers}

\author[UB]{Alessandro Mossa}
\author[UB]{Josep Maria Huguet}
\author[UB,CIBER]{Felix Ritort\corref{cor}}

\address[UB]{Departament de F\'{\i}sica Fonamental, Facultat de F\'{\i}sica, Universitat de Barcelona, Diagonal 647, E-08028, Barcelona}
\address[CIBER]{CIBER de Bioingenier\'{\i}a, Biomateriales y Nanomedicina, Instituto de Salud Carlos III, Madrid}

\cortext[cor]{Corresponding author. Fax: +34-93-402-1149. \ead{ritort@ffn.ub.es}}

\begin{document}

\begin{abstract}
We present two examples of how single-molecule experimental techniques applied to biological systems can give insight into problems within the scope of equilibrium and nonequilibrium mesoscopic thermodynamics. The first example is the mapping of the free energy landscape of a macromolecule, the second the experimental verification of Crooks' fluctuation theorem. In both cases the experimental setup comprises optical tweezers and DNA molecules.
\end{abstract}

 
\maketitle

\section{Introduction} \label{sec:intro}

Mesoscopic thermodynamics (also known as thermodynamics of small systems \cite{Hill:2002aa})   deals with the intermediate scale between the microscopic and macroscopic level. At such scale the typical number of microscopic components is much larger than 1 but much smaller than Avogadro's number, the energy exchanged between the system and its environment is of the order of few $k\sub{B}T$, and fluctuations of thermodynamic quantities about their expected values are both large and meaningful, that is, they are easily detected and carry information about the structure of the system itself. 

Many of the systems investigated in biophysics are small in the sense defined above.
This is why the recent advances in single-molecule experimental techniques \cite{Ritort:2006aa} that allow the manipulation of biological nanostructures, such as proteins or nucleic acids, provide with a valuable investigation tool not only the biologist, but also the physicist interested into the mesoscale world. 

The goal of this paper is to offer a taste of the potentiality of these techniques; it is addressed primarily to physicists not already familiar with the subject. We have selected, among many active research fields, two topics we have been working on using optical tweezers in our laboratory in Barcelona. The former belongs to equilibrium thermodynamics: in Sec.~\ref{sec:FE} we discuss how to map the free energy landscape of a biological macromolecule, specifically a long DNA hairpin. The latter is a problem of nonequilibrium thermodynamics: the experimental verification of Crooks' fluctuation theorem, to be treated in Sec.~\ref{sec:FT}. Before delving into the applications, however, we sketch in the next section a brief outline of the experimental setting.

\section{Optical tweezers} \label{sec:OT}

\begin{figure}
\centering
\includegraphics[width=8.11cm]{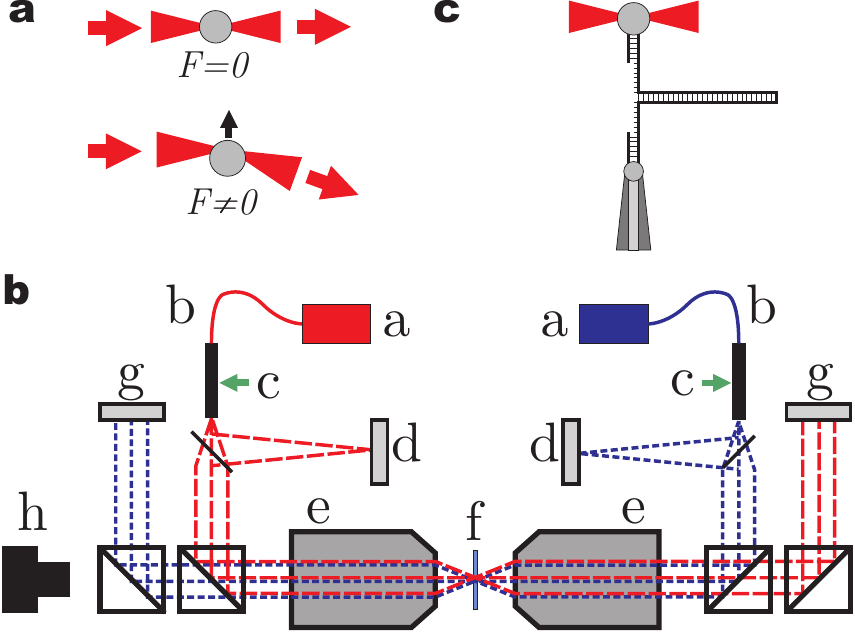} 
\caption{Optical trapping. \textbf{a} In the upper picture, a transparent bead is trapped by a focused laser beam. When the bead is centered, the incoming laser beam has the same linear momentum as the outcoming one. The bead feels no force. In the lower picture, an off-center bead undergoes a restoring force. The laser beam is deflected and the bead feels a force so that the total linear momentum is conserved. \textbf{b} Sketch of the instrument. The instrument consists on two symmetrical optical paths that produce one single optical trap. Laser beams are produced in laser diode moduli (a). The light is conducted to the so-called \emph{wigglers} (c) through an optical fiber (b). The wigglers redirect the light using piezo actuators that change the position of the optical trap. A small amount of light is split from the main beam and collected by a Position Sensitive Detector (PSD) that records the position of the optical trap (d). The remaining beam is used to form the optical trap in the fluidic chamber (f) by means of a microscope objective (e). The outcoming beam is collected by the other objective and evaluated by another PSD that measures the force (g). The optical trap and the whole experiments are observed using a CCD camera (h). \textbf{c} Experimental setup. Each strand of the DNA molecule to be unzipped is bonded to equal length dsDNA handles and these are tethered to the bead. The upper bead follows the optical trap when it is moved upwards and the force transmitted along the handles produces the unzipping of DNA \cite{Essevaz-Roulet:1997, Rief:1999aa, Bockelman:2002aa, Danilowicz:2003aa}}
\label{fig:OT}
\end{figure}

Optical tweezers \cite{Ashkin:1997aa} use the pressure of light radiation to measure and apply force to small transparent objects, such as microscopic spherical particles (usually called \emph{beads}) made of polystyrene, latex or silica. A bead illuminated by a laser beam undergoes two forces: one is proportional to the gradient of the intensity of light, the other is the scattering force due to the light reflected on the bead surface. If the gradient and scattering force are equilibrated, an optical trap is formed. This can be achieved by using a well focused Gaussian beam with high numerical aperture. The measurement of the deflection of the laser beam due to its interaction with the bead makes possible the evaluation of the applied force, once the principle of conservation of light momentum is taken into account (see Fig.~\ref{fig:OT}a).

Our apparatus (see Fig.~\ref{fig:OT}b) is a dual counterpropagating beam miniaturized optical tweezers \cite{Patent}, with
fiber-coupled diode-lasers that produce a piezo controlled movable optical trap. It is possible to apply and measure forces in the range $1\div100$ pN. The instrument also measures variations of the position of the optical trap's center. The resolution is 0.1 pN in force and 0.5 nm in position, and the sampling rate is 1 kHz. The experiments are performed in a fluidic chamber so that the molecule of interest is surrounded by water buffer. A thoroughly detailed description of our experimental apparatus is due to appear in a forthcoming paper \cite{Huguet}, while an effective theoretical model for the characterization of the bead in the optical trap and the handles that hold the molecule can be found in Ref.~\cite{Manosas:2005aa}.

\begin{figure}
\centering
\includegraphics[width=8.0cm]{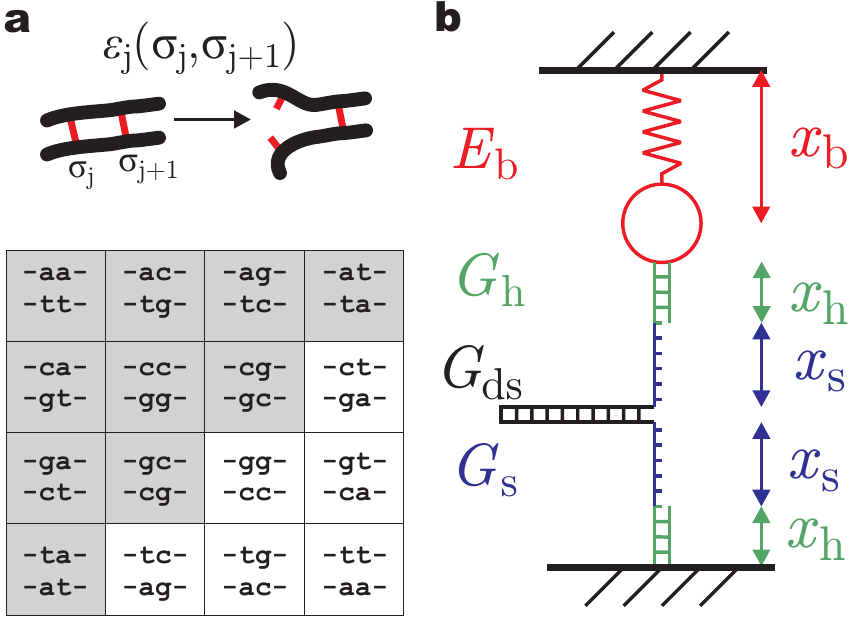}
\caption{Nearest-Neighbor model. \textbf{a} The formation energy of the molecule depends on the formation energy of each single base-pair. Since there are 4 types of bases (Adenine, Guanine, Cytosine and Thymine), 16 different NN base-pair interactions are possible. However, due to symmetry reasons, they get reduced down to the 10 independent parameters highlighted in the table. \textbf{b} The description of the experiment is broken down into different elements. Each element contributes to the total distance of the system according to its extension at a given force. The total energy of the system is the sum of the contributions of the separate elements.}
 \label{fig:NNmodel}
\end{figure}

The DNA molecule is labeled with biotin and digoxigenin groups in both ends. These labels stick to antigen coated polystyrene beads (sized few micrometers). A pulling experiment is set when one bead is in the trap and the other is held in a fixed position by a micropipette (see Fig.~\ref{fig:OT}c). A canonical experiment consists of repeated pulling and relaxing cycles during which the two beads are periodically separated and brought back together. The rapidity of the process is characterized either as pulling speed (expressed in nm/s) or as \emph{loading rate} (measured in pN/s). Data are collected in the form of force-distance curves (see, for example, Fig.~\ref{fig:traj}), from which a wealth of information about the molecule can be mined.

\section{Mapping the free energy landscape of DNA molecules} \label{sec:FE}

Using the apparatus described in the previous section, we have performed mechanical unzipping experiments on long (thousands of base-pairs) DNA molecules \cite{Huguet}. These experiments are well described by the Nearest-Neighbor (NN) model for nucleic acids \cite{Santalucia:1998aa}, which predicts the free energy of formation of a dsDNA (double-stranded DNA) from two complementary sequences of ssDNA (single-stranded DNA). According to the NN model, the free energy of formation of the $j$-th base-pair $\varepsilon_j$ depends not only on the bases that appear in position $j$, but also on the nearest-neighbor base-pair in position $j+1$ (see Fig.~\ref{fig:NNmodel}a). The total free energy of formation of a molecule is simply given by the sum of all individual contributions\footnote{Actually, one needs to distinguish a loop contribution at the end of the chain.}
\be \label{eq:Gds}
	 G_\mathrm{ds} = \sum_j\varepsilon_j(\sigma_j,\sigma_{j+1}) \,,
\ee
where the discrete variables $\sigma_j$ take values in the set of 4 canonical couplings between the bases: 
\[ 
\sigma_j\in\{AT,TA,CG,GC\}\,.
\] 
Such free energy is explicitly sequence-dependent.

The additive nature of the free energy (\ref{eq:Gds}) allows us to calculate the free energy $G_\mathrm{ds}(n)$ of intermediate states, i.e., when the first $n$ base-pair bonds are broken. An interesting application of this model is that a DNA sequence can be specifically designed in order to obtain the desired free energy profile.

Provided that the process is in quasi-equilibrium and the force $f$ along the system is homogenous at any moment, the total free energy is simply the sum of the contributions from all the components: the trap, the handles, the double- and single-stranded DNA.
The energy $E_\mathrm{b}$ of the bead in the trap is approximately quadratic in the distance $x_\mathrm{b}$ of the bead's center from the trap's focus:
\be
	E\sub{b}=\frac{1}{2}k\sub{b}x\sub{b}^2=\frac{f^2}{2k\sub{b}} \,,
\ee
where $k\sub{b}$ is the stiffness of the trap, function of the power of the laser and the properties of the bead (shape, size, and material), and the latter equality follows from $f=k\sub{b}x\sub{b}$.
 The elastic free energies of the polymers are calculated by integration of the appropriate force vs.\ extension curves:
\begin{equation}\label{eq:gdei}
G\sub{i}(x\sub{i})=\int_0^{x\sub{i}} F\sub{i}(x)\, \mathrm{d}x
\end{equation}
where $i$ is either $s$ for the ssDNA  (modeled by a Stretched Freely-Jointed-Chain \cite{Smith:1996aa}) or $h$ for the handles (modeled by a Worm-Like-Chain \cite{Bustamante:1994aa}).

\begin{figure}
\includegraphics[width=8.8cm]{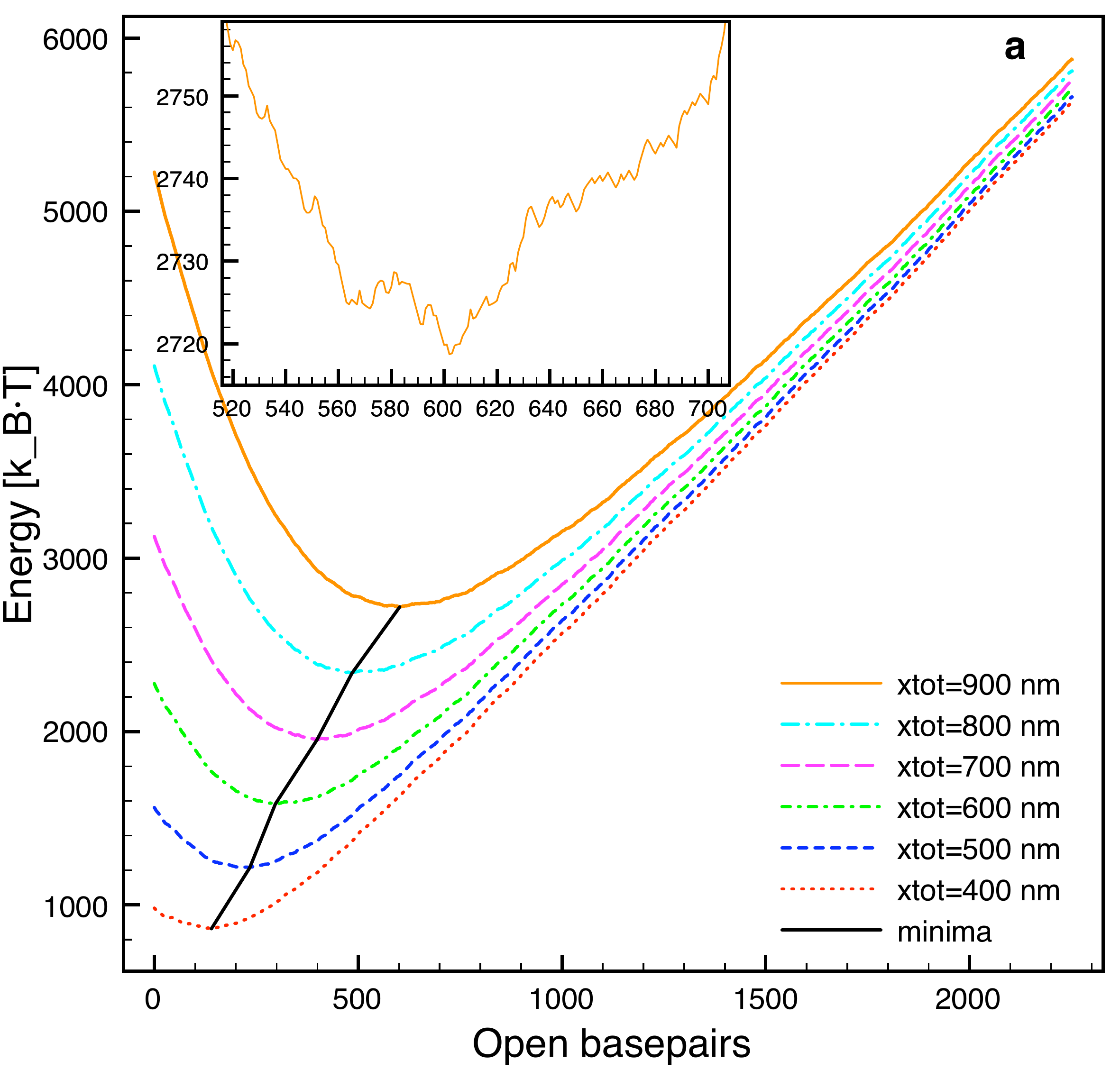}
\caption{Free energy landscape of a 2252 base-pair (bp) molecule at fixed $x\sub{tot}$. \textbf{a} Several free energy vs.\ number of open base-pairs curves are plotted for different total distances. At a fixed total distance, the energy landscape shows a minimum of energy which is assumed to be the most probable state. The inset shows a detailed view where the roughness of the free energy landscape can be appreciated. The global minimum is surrounded by other local minima that may coexist with it.} \label{fig:xtotfel}
\end{figure}

The explicit calculation depends on the statistical ensemble, which is
determined by the control parameter of the unzipping experiments. For
optical tweezers, the control parameter can be either the total
extension $x_\mathrm{tot}$ (equal to
$x_\mathrm{b}+2x_\mathrm{h}+2x_\mathrm{s}$, see
Fig.~\ref{fig:NNmodel}b) or the applied force\footnote{Magnetic tweezers are ideal setups to control force \cite{Danilowicz:2003aa}. Also specifically designed tweezers setups with zero
    stiffness regions \cite{Greenleaf:2005aa} can operate at controlled
    force. Force-feedback systems are not ideal constant force systems
    as they introduce other sort of noise effects due to the limited
    feedback frequency (typically around 1 kHz).} $f$. In the former case
the free energy is:

\be \label{eq:xtotfreee}
	G(x_\mathrm{tot},n)=E_\mathrm{b}(x_\mathrm{b})+2G_\mathrm{h}(x_\mathrm{h})+2 G_\mathrm{s}(x_\mathrm{s},n)+G_\mathrm{ds}(n) \,.
\ee

As for the constant force ensemble, the calculation of the total free energy $H(f,n)$ comprises the same elements that appear in Eq.~(\ref{eq:xtotfreee}):
\be \label{eq:ffreee}
	H(f,n)=E\sub{b}(f)+2H\sub{h}(f)+2H\sub{s}(f,n)+G\sub{ds}(n) \,,
\ee
where $E\sub{b}$ is written in terms of the applied force and the free energy of the polymers is given by the Legendre transform of Eq.~(\ref{eq:gdei}):
\begin{equation}
H\sub{i}(f)=-\int_0^{f}X\sub{i}(f')\,\mathrm{d}f' \,,
\end{equation}
where $i$ stands for either $s$ or $h$ and $X\sub{i}(f)$ is the inverse of $F\sub{i}(x)$.

Equations (\ref{eq:xtotfreee}) and (\ref{eq:ffreee}) are the basis for the study of the equilibrium thermodynamics of DNA unzipping. Several important properties can be investigated; here we focus on the reconstruction of free energy landscapes.

\begin{figure}
\includegraphics[width=8.8cm]{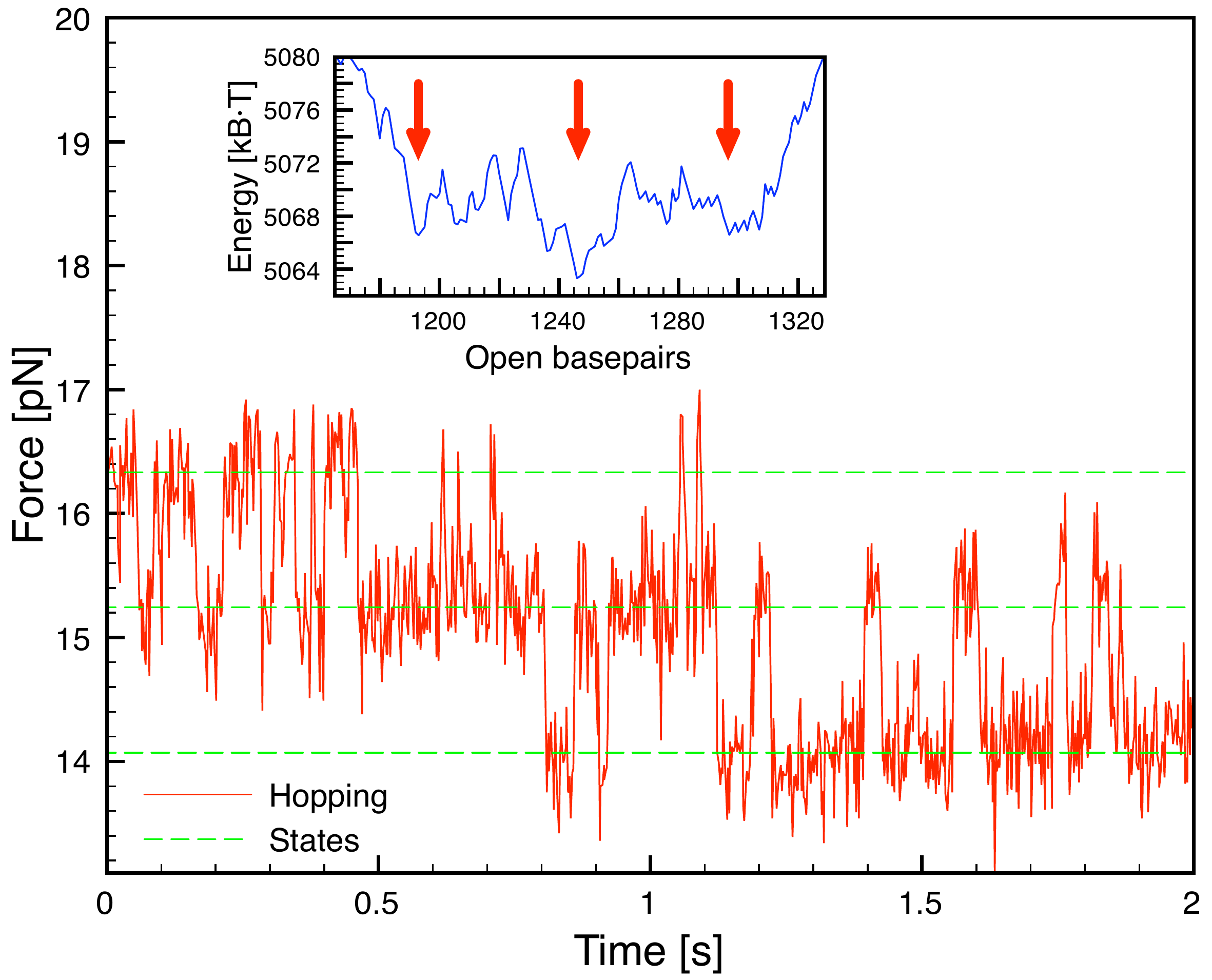}
\caption{Experimental data of hopping transitions between 3 basins that coexist. Although the effects of some intermediate states are detected, they can not be identified because their transitions are masked by thermal fluctuations. The theoretical calculation of the free energy landscape allows us to recognize  3  minima or basins that are separated by barriers that are in the range  $5 \div 10\ k\sub{B}T$, as shown in the inset. The free energy difference among them is small enough to experimentally observe the coexistence.} \label{fig:coexistence}
\end{figure}

Figure \ref{fig:xtotfel} shows a plot of the numerically computed \cite{Zuker:2003aa} free energy landscape at fixed total distance. The tuning of the control parameter $x\sub{tot}$ changes the shape of the landscape, and the stability of the states characterized by $n$ is being modified. In a typical unzipping experiment, where $x\sub{tot}$ is slowly increased at a fixed rate, the DNA molecule passes through a succession of states characterized by the number of broken bonds $n$. The unzipping is not a correlative process in which each base-pair bond is broken before the next one. Instead, it is a cooperative phenomenon in which the base-pair bonds are broken in groups. This agrees with a rough landscape where minima in the free energy abruptly change from state $n_1$ to $n_2$ (where $n_2-n_1$ is usually larger than 1) as $x_\mathrm{tot}$ increases.

The experiments also show coexistence of states in which hopping between them can be observed. The residence time of the states is about 1 s, so the time resolution of the instrument (1 ms) allows us to identify this process. This phenomenon is well described by the roughness of the free energy landscape, too. Figure \ref{fig:coexistence} shows a magnified region of the free energy landscape when the coexistence of 3 states is observed for a finely tuned value of $x\sub{tot}$.

\begin{figure}
\includegraphics[width=8.8cm]{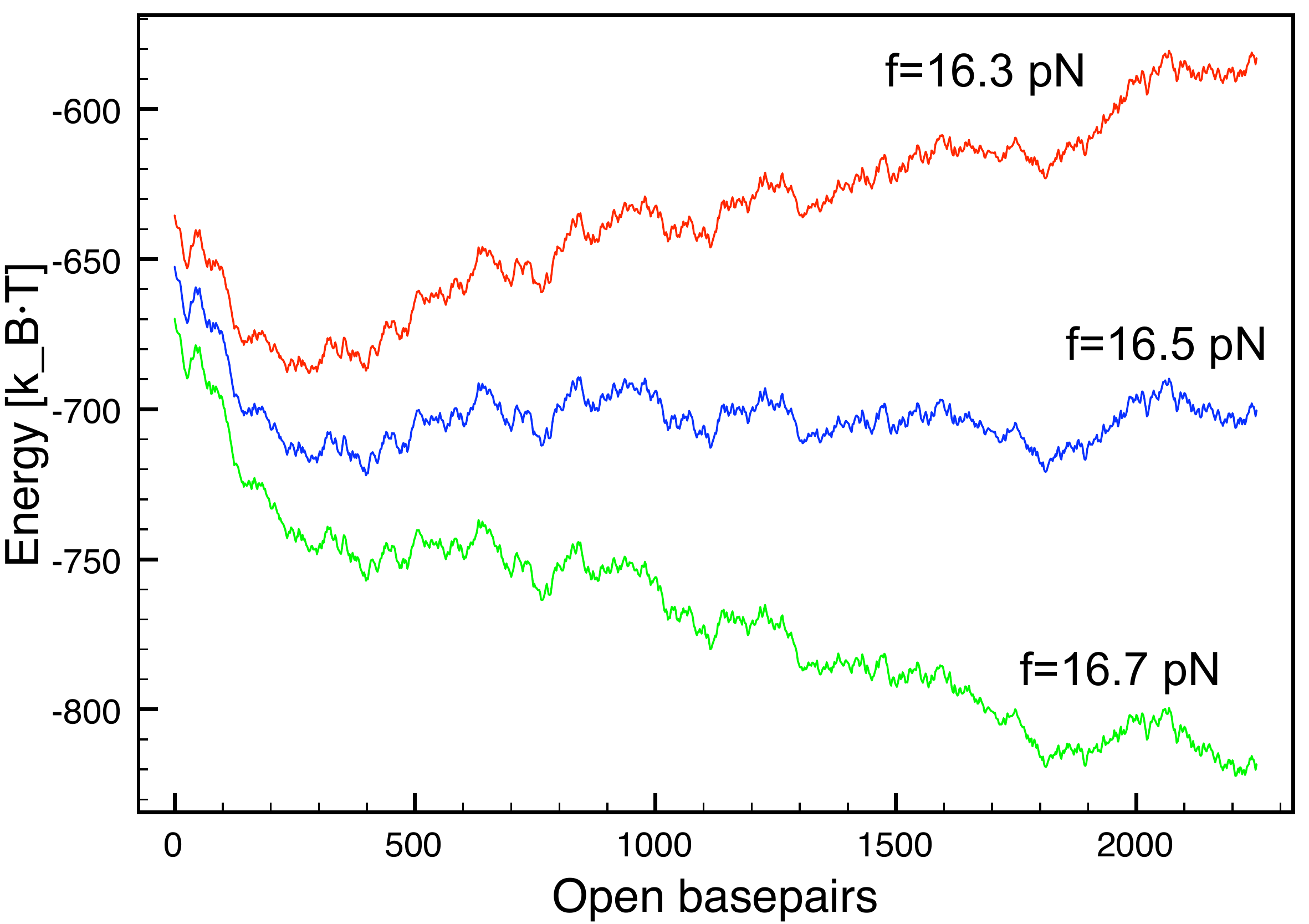}
\caption{Free energy landscape of a 2252 base-pair (bp) molecule at fixed $f$. Three free energy vs.\ number of open base-pairs curves are plotted for different forces which induce different tilts of the landscape. There is a coexistence force ($f=16.5$ pN) at which the mean tilt is zero. In this situation, many states with different number of open base pairs coexist.} \label{fig:ffel}
\end{figure}

As can be seen in Fig.~\ref{fig:ffel}, similar but slightly different aspects and details are emphasized by the free energy landscape of the system at fixed force. In this case, the force induces a tilt in the free energy landscape. From this description, we can deduce that there is a \emph{coexistence} force that induces no tilt in the force landscape. Under this situation, there are a lot of minima along the molecule that coexist and may be observed. If the force is increased, the tilt is induced, the states with higher number of broken bonds become more stable and the molecule begins unzipping. On the contrary, if the force is reduced the molecule is re-zipped. Interestingly, the tilt can be adjusted by tuning the constant force applied on the molecule.

The features of the system we are exploring allow us to design experiments to study the diffusion in an unidimensional rough landscape. A particular sequence of DNA can also be synthesized in order to produce desired peaks and valleys. The combination of the experimental measurements and the whole theoretical description has been fruitful and allows us to explore the equilibrium and close-to-equilibrium properties of DNA base-pair interactions.

\section{Experimental test of Crooks' fluctuation theorem} \label{sec:FT}

The understanding of nonequilibrium thermodynamics of small systems \cite{Bustamante:2005aa} experienced in the last ten years a season of rapid progress, whose milestones are the theoretical results collectively known as fluctuation theorems \cite{Evans:2002aa} and their experimental verifications. Here we are concerned with one such theorem, due to Crooks \cite{Crooks:1999aa}. 

Let us consider a small system immersed in a thermic bath at temperature $T$ (for instance, our setup of optical trap + beads + handles + DNA molecule). Let us say we can manipulate the system by varying the value of a control parameter $\lambda$ (for us, the total distance $x\sub{tot}$ between the trap and the pipette). Other parameters that identify the state of the system and cannot be affected directly, we group them together in a variable $s(\lambda)$. In our example, $s$ may be a discrete variable assuming value 0 if the DNA hairpin is closed and 1 if it is open\footnote{Here we are considering a DNA molecule short enough that it opens all at once, without the intermediate states considered in section \ref{sec:FE}.}. If the system is in thermodynamic equilibrium, then the probability density of $s$ follows the Maxwell--Boltzmann distribution. Now we choose an experimental protocol $\lambda\sub{F}(t)$, which we label as `forward', that starts in equilibrium at time $t=0$ with $\lambda=\lambda_0$ and ends out of equilibrium at time $t = \tau$ with $\lambda=\lambda_\tau$. Another protocol, labeled `reverse', also starts from equilibrium and is such that $\lambda\sub{R}(t)=\lambda\sub{F}(\tau-t)$. 

\begin{figure}
\includegraphics[width=8.8cm]{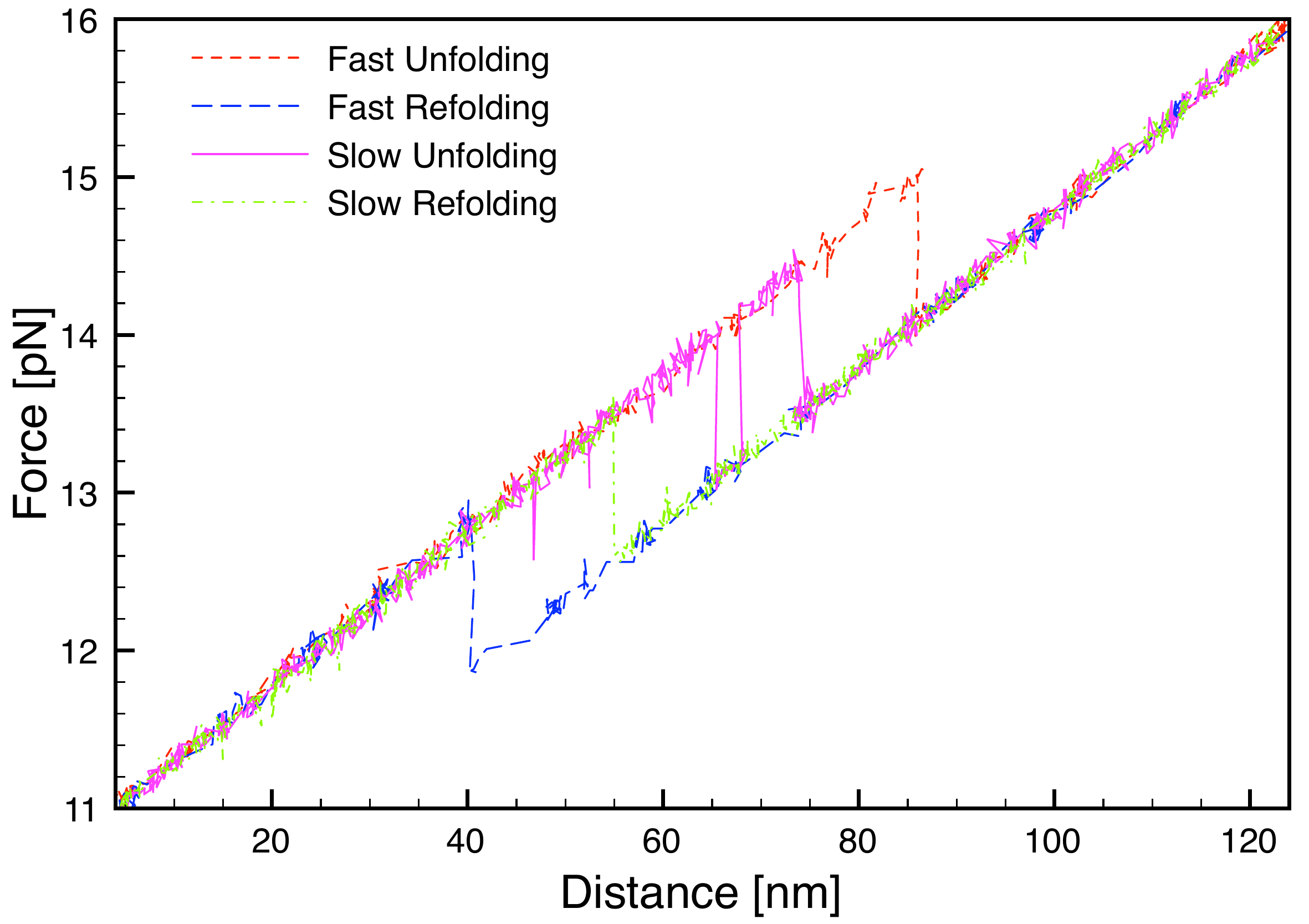}
\caption{Some representative force-distance curves. Here the ``fast'' cycle of unfolding/refolding is carried on at 400 nm/s (corresponding to a loading rate of 18.4 pN/s), and the ``slow'' one at 40 nm/s (or 1.84 pN/s).} \label{fig:traj}
\end{figure}

Let the work $W$ be defined as the energy that we feed into the system (the whole experimental system, trap included) throughout the nonequilibrium process of varying $\lambda$.
We indicate with $\mathcal{P}\sub{F}(W)$ and $\mathcal{P}\sub{R}(W)$ the work probability densities along the forward and reverse process, respectively. Crook's theorem states that, provided that the microscopic dynamics always obeys the detailed balance condition \cite{Crooks:2000aa}, 
\be \label{eq:Crooks}
	\frac{\mathcal{P}\sub{F}(W)}{\mathcal{P}\sub{R}(-W)}=\exp\left(\frac{W-\Delta G}{k\sub{B}T}\right) \,,
\ee
where $k\sub{B}$ is the Boltzmann constant, and $\Delta G$ is the equilibrium free energy difference 
\be
\Delta G\equiv G(\lambda_\tau,s(\lambda_\tau))-G(\lambda_0,s(\lambda_0))\,.
\ee

\begin{figure}
\includegraphics[width=8.8cm]{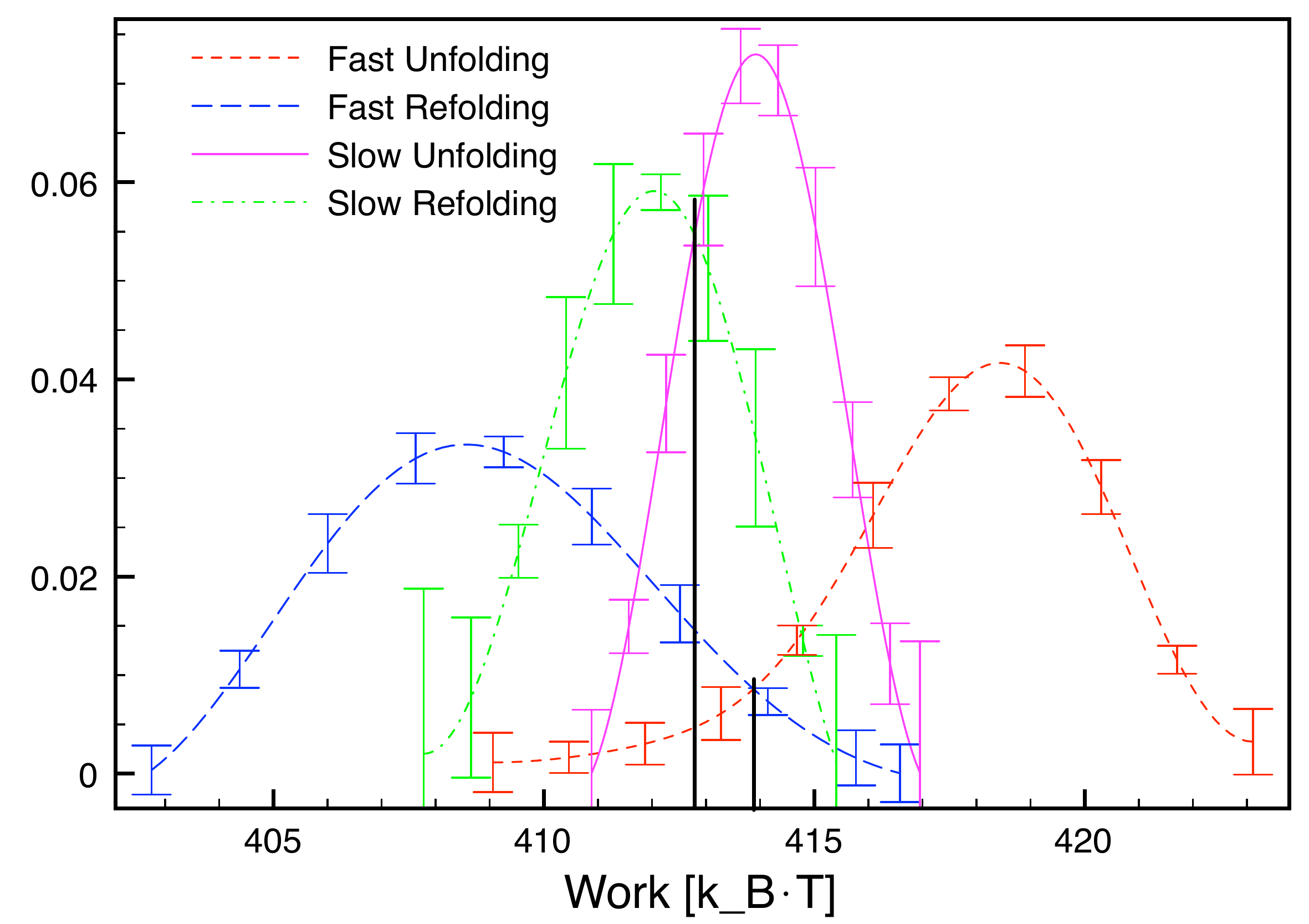}
\caption{Work distributions $\mathcal{P}\sub{F}(W)$ (unfolding) and $\mathcal{P}\sub{R}(-W)$ (refolding). The vertical lines highlight the crossing points. The distance between these two estimates of the free energy is about 1.1 $k\sub{B}T$. The distributions are evaluated from 561 fast and 228 slow unfolding/refolding cycles.} \label{fig:workdistrib}
\end{figure}

From Eq.~(\ref{eq:Crooks}), with elementary manipulations, we get
\be
	\exp(-\beta\Delta G)\mathcal{P}\sub{R}(-W) = \mathcal{P}\sub{F}(W)\exp(-\beta W) \,,
\ee
where $\beta$ is as usual the inverse of $k\sub{B}T$.
Now we can multiply both sides by a generic function $\phi(W)$ and integrate over $W$. The result can be arranged in the form
\be
	\label{eq:Crooks2}
	\exp(-\beta\Delta G)=\frac{\langle\phi(W)\exp(-\beta W)\rangle\sub{F}}{\langle\phi(-W)\rangle\sub{R}} \,,
\ee
where the angular brackets $\langle\cdots\rangle_{\mathrm{F}(\mathrm{R})}$ stand for an average over all possible realizations of the forward (reverse) protocol. The simplest possible choice, $\phi(W)=1$, yields the well-known Jarzynski identity \cite{Jarzynski:1997aa}. To the purpose of measuring free energies, however, there is a more convenient option.

In practice, one estimates the averages that appear in Eq.~(\ref{eq:Crooks2}) using a finite number of experimental events. Equation (\ref{eq:Crooks2}) can therefore be interpreted as the definition of an estimator of $\Delta G$. It turns out that the statistical variance of such estimator is minimized by Bennett's function \cite{Bennett:1976aa}
\be \label{eq:Bennett}
	\phi(W)=\left\{1+\frac{n\sub{F}}{n\sub{R}}\exp[\beta(W-\Delta G)]\right\}^{-1} \,,
\ee
where $n\sub{F}$, $n\sub{R}$ is the number of forward or reverse events, respectively.

Following the steps of the first experimental test of Crooks' theorem \cite{Collin:2005aa}, we have performed pulling experiments with optical tweezers on a short (20 bp) DNA hairpin, characterized by a two-state behaviour \cite{Ritort:2004aa}. In general, when one is interested in studying some physical property of a particular molecule, it is advisable to collect data from as many specimens as possible. Here, however, our goal is to illustrate the validity of Eq.~(\ref{eq:Crooks}), and the sake of clarity would be poorly served by the variability brought about by the inevitable individual differences in a large sample. We use therefore data taken at two different pulling speed (fast, 400 nm/s, and slow, 40 nm/s), but from the same specimen. In this way we are assured that we are always working with the same object, and can concentrate on the issue of how precisely the Crooks fluctuation relation can be checked in actual experiments. 

\begin{figure}
\includegraphics[width=8.8cm]{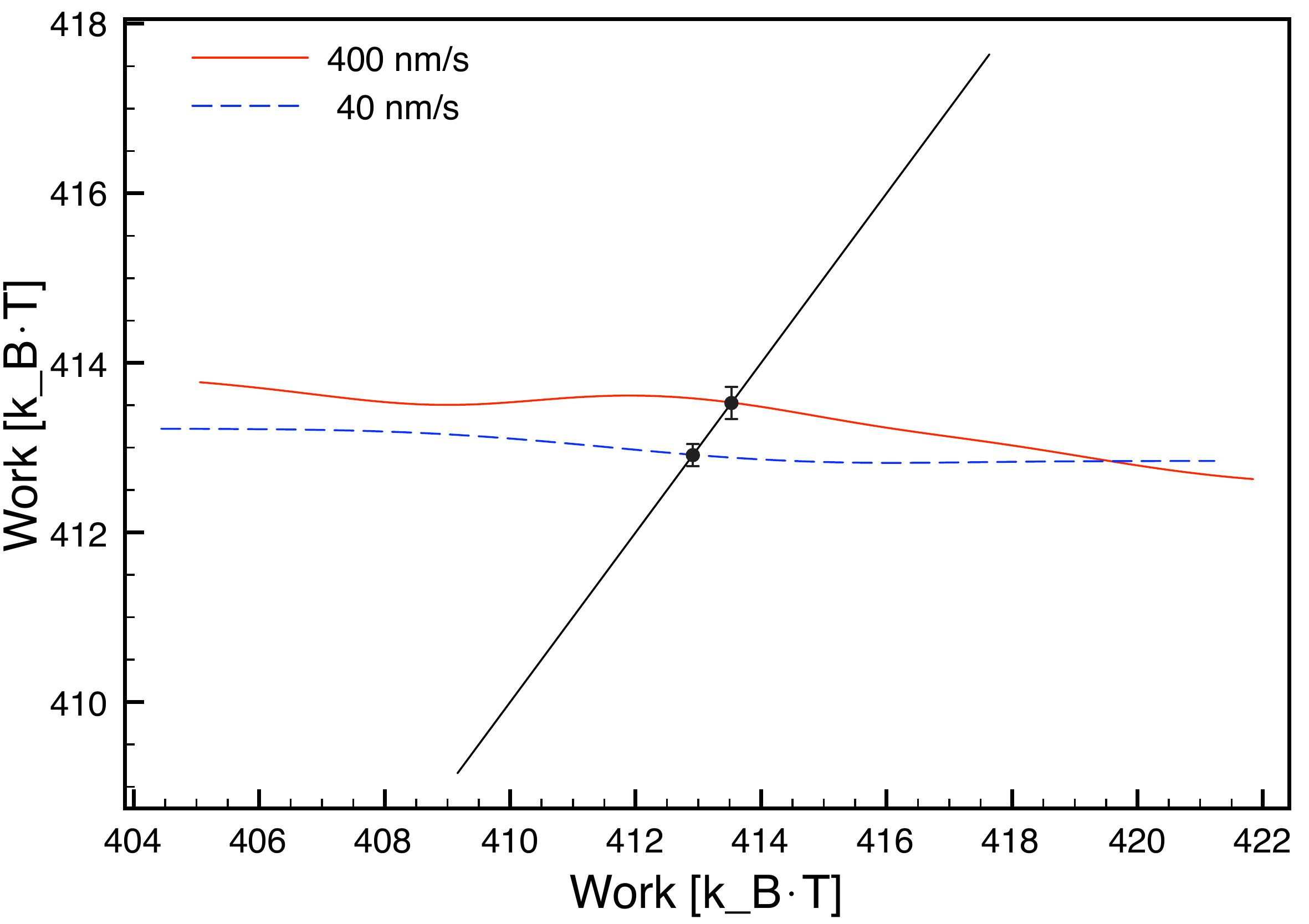}
\caption{Bennett's so-called ``acceptance ratio'' method. The nearly horizontal lines are the functions $z\sub{R}(u)-z\sub{F}(u)$ for fast and slow trajectories. They intersect the line $y=u$ at 413.53(20) $k\sub{B}T$ and 412.91(14) $k\sub{B}T$, respectively.} \label{fig:bennett}
\end{figure}

In Fig.~\ref{fig:traj} we show few typical force-distance curves $f(\lambda)$.  
The 1 pN jump in the force is the signal of the unfolding or refolding transition. It may be observed how the fast cycle exhibit larger hysteresis, while in the slow process it is not infrequent to have multiple transitions between the closed and the open states. As the pulling speed is reduced, the so-called \emph{transient violations} of the second law \cite{Ritort:2003aa}, that is, cycles in which the energy expense is negative, also become more probable.

From the force-distance curves we can compute the mechanical work performed on the system
\be
	W=\int_{\lambda_0}^{\lambda_\tau} f(\lambda)\,\rmd\lambda \,.
\ee 
Figure \ref{fig:workdistrib} shows the estimated work distributions thus obtained. The work probability density functions have been evaluated according to a histogram-free method recently proposed in Ref.~\cite{Berg:2008aa}: one nice advantage over traditional, histogram-based methods is that the theoretically expected asymmetry of the work distribution \cite{Saha:2007aa} is evident, at least in the ``fast'' case.

Equation (\ref{eq:Crooks}) implies that the graphs of $\mathcal{P}\sub{F}(W)$ and $\mathcal{P}\sub{R}(-W)$ cross each other for $W=\Delta G$: the work distributions themselves are dependent on the pulling speed, but the crossing point is not. In our example, the distance between the crossing point of the fast distributions is about 1.1 $k\sub{B}T\simeq 0.6$ kcal/mol larger than the crossing point of the slow ones. However, we should observe that the errors on the ``slow'' distributions are large, besides, this is hardly the best way of estimating $\Delta G$. As explained earlier, the most convenient way to extract a free energy estimate from Eq.~(\ref{eq:Crooks}) is to use Bennett's choice (\ref{eq:Bennett}) into Eq.~(\ref{eq:Crooks2}).

In practice, we define the functions
\begin{subequations}
\begin{gather}
	z\sub{F}(u)  = \ln \left\langle\frac{\exp(-\beta W)}{1+\exp[\beta(W-u)]}\right\rangle\sub{F} \,, \\
	z\sub{R}(u)  = \ln \left\langle\frac{1}{1+\exp[-\beta(W+u)]}\right\rangle\sub{R} \,,
\end{gather}
\end{subequations}
where we are taking advantage of the fact that for us $n\sub{F}=n\sub{R}$, and plot the function, that should be approximately constant, $z\sub{R}(u)-z\sub{F}(u)$ (see Fig.~\ref{fig:bennett}). The intersection with the line $y=u$ is the optimal estimate of $\Delta G$. To this number we can attach a standard deviation, as explained in Ref.~\cite{Shirts:2003aa}: in our example, we find $\Delta G = 413.53(20)$ $k\sub{B}T$ from the ``fast'' data and $\Delta G = 412.91(14)$ $k\sub{B}T$ from the ``slow'' ones. The discrepancy between the two values has been reduced to about 0.6 $k\sub{B}T\simeq 0.4$ kcal/mol, and is consistent with the hypothesis that they are in fact two estimates of the same quantity, as required by Crooks' theorem.

\begin{figure}
\includegraphics[width=8.8cm]{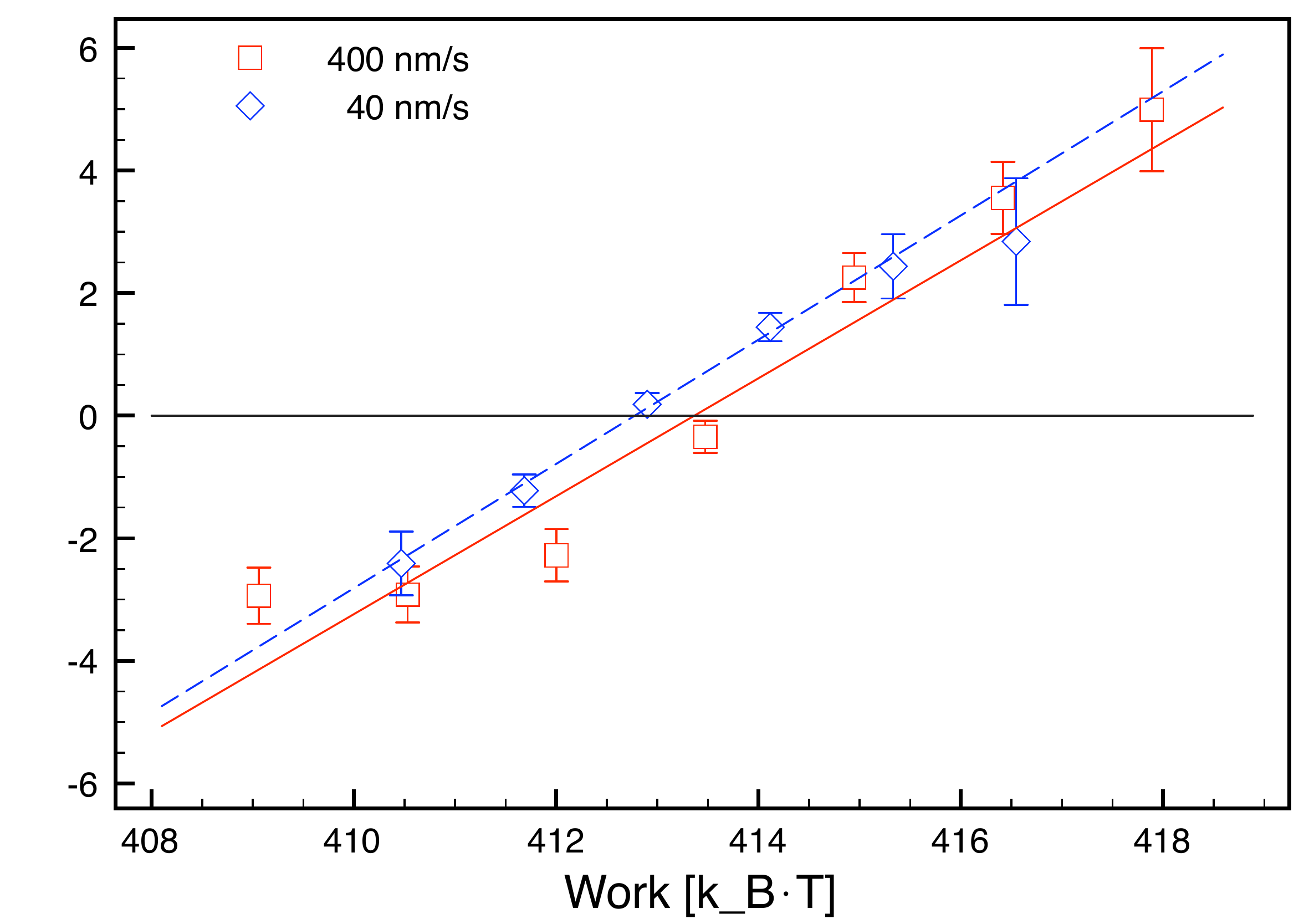}
\caption{Plot of $\ln(\mathcal{P}\sub{F}(W)/\mathcal{P}\sub{R}(-W))$ as a function of $W$. The least squares linear fit slope is $0.96(7)$ for the ``fast'' data and $1.0(1)$ for the ``slow'' ones.} \label{fig:lines}
\end{figure}

Another way of testing Eq.~(\ref{eq:Crooks}) is to directly plot the quantity $\ln(\mathcal{P}\sub{F}(W)/\mathcal{P}\sub{R}(-W))$, as in Fig.~\ref{fig:lines}. If the Crooks fluctuation relation is satisfied, then the curves must be straight lines with slope 1 which cross the $W$-axis at $\beta\Delta G$. The slopes of the linear fit that we find in this way, $0.96(7)$ and $1.0(1)$ for the fast and slow case, respectively, are fully consistent with the theoretical expectations.

\section{Conclusion} \label{sec:bye}

Single-molecule experimental techniques such as the atomic force microscope and optical or magnetic tweezers are now established as a routine tool in biophysical research. In this paper we have illustrated how they can be equally useful for the physicist interested in the mesoscale phenomenology. Easily synthesized biological macromolecules are an ideal laboratory  to verify our comprehension of equilibrium and nonequilibrium thermodynamics of small systems.  

\section*{Acknowledgments}
We thank Marco Ribezzi for a critical reading of the manuscript. We acknowledge financial support from grants FIS2007-61433, NAN2004-9348, and SGR05-00688.

\end{document}